# Dose–LET Interactions Predict Capsular Contracture After Proton Postmastectomy Radiation Therapy

**Authors:** Jingyuan Chen, PhD[1,#], Zeliang Ma, MD[2,#], Meiyun Cao, MSc[1], Robert W. Gao, MD[2], Yunze Yang, PhD[3], Yuzhen Ding, PhD[1], Nicholas B. Remmes, PhD[2], Jiasen Ma, PhD[2], Kimberly S. Corbin. MD[2], Dean A. Shumway, MD[2], Robert W. Mutter, MD[2,*] and Wei Liu, PhD[1,*]

[1]Department of Radiation Oncology, Mayo Clinic, Phoenix, AZ 85054, USA

[2]Department of Radiation Oncology, Mayo Clinic, Rochester, MN, 55905, USA

[3]Department of Radiation Oncology, the University of Miami, FL 33136, USA

[#]Co-first author

Corresponding author:

Robert W. Mutter, MD, Department of Radiation Oncology and Department of Molecular Pharmacology and Experimental Therapeutics, Mayo Clinic, 200 1st St. SW, Rochester,MN 55905, USA. Email: Mutter.Robert@mayo.edu.

Wei Liu, PhD, Department of Radiation Oncology, Mayo Clinic Arizona, 5777 E. Mayo Boulevard, Phoenix, AZ 85054, USA. E-mail: Liu.Wei@mayo.edu.

**Conflicts of Interest Disclosure Statement**

No

**Funding Statement**

This work was supported in part by the Department of Radiation Oncology, Mayo Clinic, Rochester, Minnesota, and Phoenix, Arizona, USA, R01 CA 261932 and R01 CA 272602 to Robert W. Mutter, and P30 CA015083 to MCCCC. This research was also supported by NIH/BIBIB R01EB293388, by NIH/NCI R01CA280134, by the Eric & Wendy Schmidt Fund for AI Research & Innovation, and by the Kemper Marley Foundation to Wei Liu.

**Ethical Approval**

This study was approved by Mayo Clinic Arizona institutional review board (IRB#: 24-011106).

**Data Availability Statement**

The data analyzed during the current study are not publicly available due to patient privacy concerns and institutional policies regarding protected health information (PHI). However, de-identified data that support the findings of this study are available from the corresponding author upon reasonable request and with appropriate institutional review board (IRB) approval.

**Abstract**

**Purpose:** Pencil beam scanning (PBS) proton therapy provides highly conformal dose distributions that are increasingly leveraged for postmastectomy radiation therapy (PMRT) to reduce cardiopulmonary exposure. However, implant-based reconstruction (IBR) in the setting of PMRT remains vulnerable to capsular contracture, and biological mechanisms of possible high linear energy transfer (LET) in PBS have not been well characterized. This study aimed to investigate the combined effects of dose and dose-averaged linear-energy-transfer ($LET_d$) on capsular contracture after proton PMRT and to derive clinically interpretable dose–LET volume constraints (DLVCs).

**Methods**: A retrospective case-control study was conducted on consecutive breast cancer patients who underwent mastectomy followed by implant-based reconstruction and proton PMRT (50 Gy in 25 fractions) between 2015 and 2021. Patients with capsular contracture were matched 1:2 with controls using nearest-neighbor matching based on clinical and pathological variables. Dose-LET volume histograms (DLVHs) were calculated for peri-implant tissue (5-mm shell around the implant). Generalized linear mixed-effects regression (GLMER) was employed to identify DLVH indices significantly associated with capsular contracture. Spearman correlation analysis was used to eliminate redundant DLVH indices. DLVCs were derived from receiver operating characteristic (ROC) analysis and validated using a support vector machine (SVM)-based normal tissue complication probability (NTCP) model with leave-one-out cross-validation.

**Results:** Eight capsular contracture patients and 16 matched controls patients were analyzed. Three independent DLVH indices were significantly associated with capsular contracture ($p<0.01$): V(55.8 Gy, 2.2 keV/μm), V(50.3 Gy, 5.4 keV/μm), and V(32.8 Gy, 0.9 keV/μm). The corresponding DLVCs were: V(55.8 Gy, 2.2 keV/μm) < 0.0033%, V(50.3 Gy, 5.4 keV/μm) < 0.0017%, and V(32.8 Gy, 0.9 keV/μm) > 96.98%. The SVM-based NTCP model achieved an area under the ROC curve (AUROC) of 0.867, with accuracy of 91.7%, sensitivity of 87.5%, and specificity of 93.8%.

**Conclusion**: Capsular contracture following proton PMRT is significantly associated with the synergistic interplay between dose and $LET_d$ in peri-implant tissue. The derived DLVCs provide actionable dosimetric constraints that can be integrated into treatment planning to minimize capsular contracture risk in breast cancer patients undergoing proton PMRT with implant-based reconstruction.

# Introduction

Pencil beam scanning (PBS) technology represents a major advancement in proton therapy, delivering significantly better dose conformity than traditional passive scattering technology that utilized apertures and compensators, especially in the proximal region of the beam path(1-9). This improved precision with PBS allows for enhanced protection of organs at risk (OARs), with skin sparing being particularly beneficial for radiotherapy treatment planning in breast cancer patients(10,11). Given these dosimetric advantages, (10-12)proton therapy has emerged as a promising modality for breast cancer because of its ability to reduce radiation exposure to the heart, lungs, and other surrounding healthy tissues compared to photon-based techniques(13,14).

For breast cancer patients, proton postmastectomy radiation therapy (PMRT) improves oncologic outcomes but can increase the risk of reconstructive complications, including infection, wound breakdown, capsular contracture, implant exposure, which causes reconstruction failure and requiring major reoperation or explantation (15,16). Capsular contracture is a common and challenging complication after postmastectomy radiotherapy (PMRT) in implant-based breast reconstruction, with incidence rates of clinically relevant contracture (Baker grade III-IV) reported around 22.9% and overall contracture (Baker I-IV) up to 47.5% (17). PMRT significantly increases the risk of capsular contracture compared to non-radiated patients and is also associated with higher rates of reconstruction failure and overall complications(18,19). PBS offers superior conformal dose distribution; however, its biological impact on periprosthetic tissue remains under-investigated. While some studies have suggested similar overall reconstruction outcomes, proton PMRT has been linked to higher rates of capsular contracture compared to photon therapy, despite similar local control (20-23).

Protons release the majority of their energy within a narrow region at the end of their trajectory, resulting in elevated linear energy transfer (LET) at the distal edge of the beam. Standard proton therapy protocols currently apply a constant relative biological effectiveness (RBE) value of 1.1 (24-27), disregarding the potential influence of spatial variations of LET distributions. Evidence of RBE values exceeding 1.1 for adverse events linked to higher LET within OAR has been documented across multiple conditions, including rib fractures (28,29), sacral fraction (30), late-phase pulmonary complications(31), brain necrosis (32-36), mandibular osteoradionecrosis (25) and rectal bleeding (37). To optimize clinical outcomes in proton therapy, it is crucial to better understand the combined influence of physical dose and LET on adverse events.

Facing the challenge of substantial biological and parametric uncertainties in current RBE models(39-42), we adopt the Dose-LET volume histogram (DLVH) proposed recently (37,43). The dose-LET volume histogram (DLVH) has emerged as an innovative approach that integrates the combined effects of dose and $LET_d$ when analyzing patient outcomes(30,43,44). By relying directly on dose and $LET_d$ rather than RBE, DLVH effectively circumvents the substantial uncertainties associated with current RBE models. This tool has demonstrated its value in studies examining adverse events such as rectal bleeding (37), mandibular osteoradionecrosis (36), and rib fractures (29). In this study we aimed to investigate the effects of dose and $LET_d$ on capsular contracture in breast cancer patients treated with proton PMRT using DLVH-based statistical approaches.

Table 1. Patient Characteristics.

|  |  | Contracture | Control | p | SMD |
|---|---|---|---|---|---|
| n |  | 8 | 16 |  |  |
| Age |  | 46.38 (10.77) | 48.44 (10.89) | 0.665 | 0.190 |
| BMI |  | 26.39 (4.40) | 26.32 (3.72) | 0.967 | 0.018 |
| Smoking | No | 6 (75.0) | 12 (75.0) | 1.000 | <0.001 |
|  | Yes | 2 (25.0) | 4 (25.0) |  |  |
| Menopausal | Pre | 5 (62.5) | 10 (62.5) | 0.755 | 0.376 |
|  | Peri | 0 (0.0) | 1 (6.2) |  |  |
|  | Post | 3 (37.5) | 5 (31.2) |  |  |
| Side | Left | 5 (62.5) | 10 (62.5) | 1.000 | <0.001 |
|  | Right | 3 (37.5) | 6 (37.5) |  |  |
| Histology | Invasive Ductal | 7 (87.5) | 11 (68.8) | 0.503 | 0.590 |
|  | Invasive Lobular | 0 (0.0) | 2 (12.5) |  |  |
|  | Invasive Mammary | 1 (12.5) | 3 (18.8) |  |  |
| PR | No | 3 (37.5) | 6 (37.5) | 1.000 | <0.001 |
|  | Yes | 5 (62.5) | 10 (62.5) |  |  |
| Mastectomy_Type | Nipple Sparing | 3 (37.5) | 9 (56.2) | 0.665 | 0.383 |
|  | Skin Sparing | 5 (62.5) | 7 (43.8) |  |  |
| Bilateral_Mastectomy | No | 2 (25.0) | 3 (18.8) | 1.000 | 0.152 |
|  | Yes | 6 (75.0) | 13 (81.2) |  |  |
| ALND | No | 3 (37.5) | 7 (43.8) | 1.000 | 0.128 |
|  | Yes | 5 (62.5) | 9 (56.2) |  |  |
| Grade | 1 | 0 (0.0) | 1 (6.2) | 0.565 | 0.505 |
|  | 2 | 4 (50.0) | 10 (62.5) |  |  |
|  | 3 | 4 (50.0) | 5 (31.2) |  |  |
| Associated_DCIS | No | 0 (0.0) | 7 (43.8) | 0.081 | 1.247 |
|  | Yes | 8 (100.0) | 9 (56.2) |  |  |
| LVI | No | 2 (25.0) | 7 (43.8) | 0.655 | 0.403 |
|  | Yes | 6 (75.0) | 9 (56.2) |  |  |
| pT | T1 | 4 (50.0) | 8 (50.0) | 0.522 | 0.653 |
|  | T2 | 1 (12.5) | 5 (31.2) |  |  |
|  | T3 | 1 (12.5) | 2 (12.5) |  |  |
|  | Tis | 2 (25.0) | 1 (6.2) |  |  |
| pN | N0 | 2 (25.0) | 1 (6.2) | 0.541 | 0.649 |
|  | N1 | 5 (62.5) | 11 (68.8) |  |  |
|  | N2 | 1 (12.5) | 3 (18.8) |  |  |
|  | N3 | 0 (0.0) | 1 (6.2) |  |  |

**Abbreviations:** ALND: Axillary Lymph Node Dissection; BMI: Body Mass Index; DCIS: Ductal Carcinoma in Situ; LVI: Lymphovascular Invasion; PR: Progesterone Receptor; pT: Pathological Tumor Stage; pN: Pathological Nodal Stage; SMD: Standardized Mean Difference.

## Methods and Materials

### Patient cohort

A prospectively maintained departmental database was queried to identify consecutive patients with breast cancer who underwent mastectomy followed by implant-based reconstruction (IBR) and proton post-mastectomy radiotherapy (PMRT) between 2015 and 2021. To ensure a homogenous study population, we excluded patients with prior ipsilateral breast or chest wall irradiation, recurrent disease, those receiving bilateral or hypofractionated PMRT. We also excluded patients requiring replanning at any point during their treatment course to maintain the integrity of the initial LET/dose distribution analysis. Detailed surgical and radiotherapeutic techniques have been previously validated and described elsewhere (22). Capsular contracture events were adjudicated via comprehensive chart review by well experienced radiation oncologists. To mitigate potential confounding factors, patients who developed contracture were matched 1:2 with controls using a nearest-neighbor Mahalanobis distance algorithm based on age, BMI, smoking, and menopausal status; tumor laterality, histology, grade, PR status, LVI, associated DCIS, and pathologic T/N stage; and surgical factors including mastectomy type and axillary lymph node dissection (ALND).

**Treatment planning and contouring**

The clinical target volume (CTV) encompassed the chest wall; axillary nodal levels I–III; the supraclavicular nodal basin; and the internal mammary nodes. The chest wall CTV was confined anteriorly, extending no deeper than the anterior surfaces of the ribs and intercostal musculature (i.e., the ribs were excluded). For patients who underwent reconstruction, the CTV also included the entire tissue expander or implant (12). The prescribed dose to the chest wall and regional lymphatics was 50 Gy (RBE) delivered in 25 fractions. All patients were treated with multi-field optimized (MFO) PBS proton plans using two or three fields, consistent with previously published techniques. Plans were created in a commercial treatment planning system (Eclipse v15.1, Varian Medical Systems, Palo Alto, CA, USA) and optimized robustly to account for ±5 mm setup uncertainty and ±3% range uncertainty. For the CTV, robustness evaluation under the worst-case perturbed scenarios aimed to achieve D90% ≥ 90% (priority 1), D95% ≥ 95% (priority 2), and a near-maximum constraint of D0.01 cc ≤ 110% (priority 1). The skin was delineated as a 3-mm thick tissue layer beneath the body surface. The planning objective for chest wall skin coverage was to achieve D90% ≥90% (acceptable threshold). Dose constraints for this region were established at D1cc ≤105% (acceptable) and D1cc ≤102% (optimal). For the supraclavicular region, skin dose constraints were set at D1cc ≤90% (acceptable) and D1cc ≤80% (optimal). All plans satisfied institutional dose–volume constraints for target coverage and OAR limits(12). In addition, each plan was independently evaluated with an in-house Monte Carlo–based biologic dose model(38,45,46). When clinically feasible and at the physician's discretion, plan optimization also sought to reduce overlap of high $LET_d$ regions and elevated physical dose on the ribs and intercostal muscles at the most posterior extent of the CTV (12). Treatments were delivered using a Hitachi PROBEAT-V proton therapy system (Hitachi, Tokyo, Japan). Patients were treated under free-breathing conditions, with respiratory motion monitored using AlignRT (Vision RT, London, UK).

**Dose-LET-Volume Histogram**

To investigate the combined effect of dose and $LET_d$ to the normal tissue toxicity, we adopted the novel tool, DLVH. DLVHs were constructed with dose (in Gy) and $LET_d$ (in keV/um) as 2 axes, and the third dimension is the normalized volume(25,29,36,37,43,47-49). The dose reported here is the dose to medium.(38,45,46). The dose averaged LET to medium, $LET_d$, was defined as:

$$LET_d(x,y,z) = \frac{\Sigma_i \phi_E(x,y,z,E_j) SP^2(E_j) \Delta E_j}{\Sigma_i \phi_E(x,y,z,E_j) SP(E_j) \Delta E_j}$$

where $(x,y,z)$ indicates the voxel location, $\phi_E(x,y,z,E_j)$ is the per voxel energy spectrum of proton. $SP(E_j)$ is the unrestricted stopping power of proton. Dose and $LET_d$ distributions for the PBS plans were computed using an in-house fast Monte Carlo dose/LET calculation engine. This system has been rigorously commissioned and benchmarked, and it is routinely deployed in our clinical workflow as an independent platform for secondary dose verification, plan optimization, and biologic dose evaluation.

Similar to the definition of dose volume histogram (DVH) indices, the DLVH index, $V_{D,LET_d}(d,l) = V(D \geq d, LET_d \geq l)$, was defined as the normalized volume $V$ of the region of interest (surrounding implant tissues in this study) with a dose of at least d Gy and $LET_d$ of at least l keV/um.

For the DLVH calculation in this study, we employed 64 bins along both the dose dimension (1.09 Gy per bin) and the LET dimension (0.188 keV/mm per bin). Through a series of experiments adjusting bin numbers on both axes, we determined that a 64-bin configuration optimally balanced computational load, statistical noise, and fine feature resolution.

**Region of interest-peri-implant tissue**

Through preliminary DLVH comparisons across different regions of interest (ROIs), combined with clinical expert consultation, we identified peri-implant tissue as the ROI potentially associated with capsular contracture. The **peri-implant tissue** was defined as a 5-mm shell expanded circumferentially from the implant surface. This 5-mm margin was selected based on established morphological and histopathological data indicating that most radiation-induced capsular thickening, fibrotic remodeling, and inflammatory infiltration occurs within a few millimeters of the prosthesis interface, typically ranges from 1.5 to 3.0 mm (50). Therefore, a 5-mm shell provides a sufficiently conservative buffer to capture the entire fibrotic zone for dosimetric analysis. To ensure anatomical accuracy for LET and dose calculation, this volume was manually edited to exclude voxels extending into the lung parenchyma or the ambient air beyond the skin surface.

**Statistical analysis and DLVH indices significance**

DLVH on peri-implant tissue for all patients were calculated. $V_{D,LET}(d,l)$ were compared between the AE group and control group. Generalized linear mixed-effects regression (GLMER) model was employed to analyze matched case-control data. This approach accounts for within-set dependence induced by matching through a matched-set–specific random intercept, thereby mitigating residual heterogeneity across matched sets and reducing confounding attributable to set-level factors. In this model, the DLVH volume index was treated as the predictor, and the binary clinical outcome (presence vs. absence of capsular contracture) was treated as the response.

The model included two effect levels: (1) Fixed effects represented the systematic associations between DLVH index $V_{D,LET}(d,l)$, and clinical outcomes across the study population. Clinical variables such as prescription dose and reconstruction status (e.g., tissue expander) were controlled via the matching design, ensuring balance between the AE group and control group; consequently, these factors were not reintroduced as explicit fixed-effect covariates during single-feature screening. (2) Random effects captured residual between-set heterogeneity remaining after matching. Specifically, a random intercept was specified for each matched set (subclass), allowing the baseline log-odds of capsular contracture to vary across matched sets and thereby accounting for intra-set dependence.

The model was expressed as:

$$\text{logit}(P(\text{capsular contracture} = 1)) = \alpha + b_i + \beta V(d,l),$$

where $i$ indexed the $i$-th matched set (subclass), $\beta$ was the regression coefficient quantifying the association between the DLVH index, $V(d,l)$, and the likelihood of capsular contracture, $\alpha$ was the overall intercept, and $b_i$ represented the matched-set–specific random intercept.

The significance of the DLVH indices was assessed by the p-value of the regression coefficient $\beta$. A lower p-value suggests a more substantial impact of the DLVH index on the likelihood of capsular contracture. For each combination of dose = d and $LET_d$ = l, a separate GLMER model was established by treating the corresponding $V(d,l)$ as the fixed-effect predictor. To comprehensively interrogate outcome-related DLVH characteristics and derive potential clinical insights, all feasible DLVH indices within the prespecified dose and $LET_d$ ranges were evaluated in an iterative manner. This procedure generated a p-value map over the (dose, $LET_d$) plane. The resulting p-value map provides a transparent approach for identifying and visualizing DLVH indices most strongly associated with capsular contracture outcomes, thereby improving interpretability of dosimetric risk features across dose, $LET_d$, and volume in breast cancer patients treated with proton therapy. The GLMER was conducted using the generated lme4 package of R (version 4.1.2).

**Important DLVH indices extraction and Correlation analysis**

We extracted the significant DLVH indices by analyzing the p-value map features. The DLVH indices enclosed by low p-value contour lines (p=0.01) indicated the most statistically significant metrics and were considered as potential candidates. Because the candidate DLVH indices could be redundant and intercorrelated, it was important to find the metrics with minimum redundancy and maximum independence. Based on the DLVHs for each patient, Spearman's correlation tests were performed to examine the correlation among all 4,096 DLVH indices and the selected DLVH indices. Spearman's correlation coefficient maps for the candidate DLVH indices were constructed by displaying the calculated correlation coefficients in 2 dimensions with the x-axis as physical dose and the y-axis as LET, mirroring the format of the *p*-value map. Redundant DLVH indices were excluded, and one representative, statistically significant DLVH index was selected from each low-p-value region.

**Volume constraints and DLVC-based normal tissue complication probability modeling**

For each significant DLVH index, we directly use its volume index for the classification of capsular contracture AE group and control group. Area under the curve (AUC) of the receiving operation characteristics (ROC) curve was calculated to evaluate the performance of each DLVH. The volume constraint of the DLVH index was extracted from the volume threshold at the optimal operating point of the receiver operating characteristic (ROC) curve. Thus, the DLVC was derived. Statistical results were generated using Matlab 2019a (Math-Works, Inc, Natick, MA).

To evaluate the predictive value of the derived DLVCs, we constructed a multivariable NTCP model using a support vector machine (SVM). The classifier was trained by solving a soft-margin optimization problem:

$$\min_{\theta, c, s} \frac{1}{2} \| \theta \|^2 + \lambda \sum_{k=1}^{N} s_k \;\; \text{s.t.} \; g_k(\theta^\top x_k + c) \geq 1 - s_k \;\; and \;\; s_k \geq 0$$

where $x_k = (V_{d_1,l_1}^{(k)}, V_{d_2,l_2}^{(k)}, ...)$ denotes the DLVC feature vector for patient $k$, $\theta$ and $c$ are the model coefficients and intercept, and $s_k$ are slack variables that permit margin violations. The tuning parameter

$\lambda$ controls the trade-off between maximizing the margin and penalizing misclassification. The binary class label $g_k$ was defined as $+1$ for patients with adverse events (AE group) and $-1$ for controls. To account for class imbalance, a higher misclassification penalty was assigned to the AE group than to the control group. All DLVC features were standardized using z-score normalization before model fitting.

To obtain an NTCP estimate, the SVM decision scores were mapped to posterior probabilities by fitting a sigmoid calibration function. Model performance was evaluated using the leave-one-out cross-validation (LOOCV), and discrimination was quantified by the area under the ROC curve (AUROC).

## Results

### Patient Characteristics

We retrospectively reviewed a database of 145 patients treated with conventionally fractionated proton therapy from 2015 to 2021. Eight consecutive patients with documented contracture were identified and matched in a 1:2 ratio with 16 control patients without contracture. Baseline characteristics, including age and other matching variables, were well-balanced between the two cohorts (Table 1). The mean age was 46.38 years in the contracture group compared to 48.44 years in the control group (P = 0.665).

### Dose and LET Distributions in a Representative Patient

Figure 1 illustrates the physical dose and dose-averaged linear energy transfer (LET$_d$) distributions for a representative patient who developed capsular contracture. Although the prescribed target volume achieved conformal coverage of 50 Gy(RBE), the LET$_d$ distribution demonstrated a pronounced accumulation of high-LET$_d$ components (>5 keV/μm) within the peri-implant tissue. Notably, these elevated LET$_d$ regions spatially coincided with areas receiving moderate to high physical dose. This observation suggests that the combined spatial overlap of dose and LET$_d$, rather than dose alone, may contribute to fibrotic remodeling and subsequent capsular contracture in peri-implant tissues.

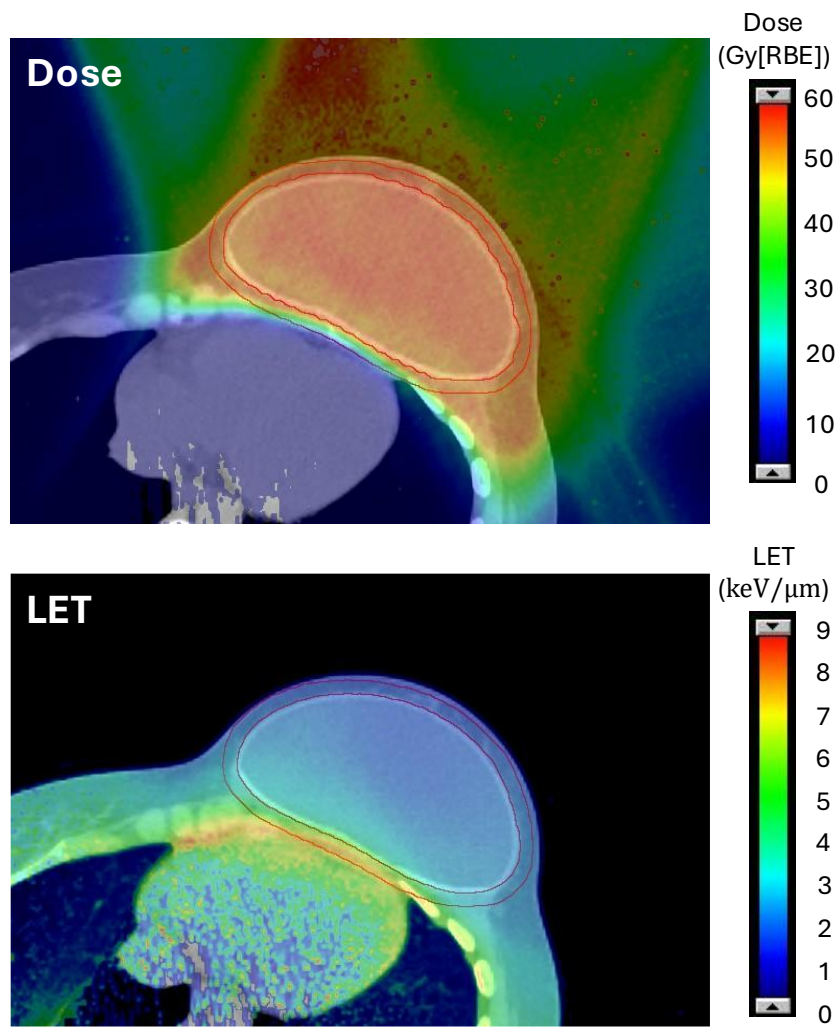

**Figure 1.** Dose (top) and LET$_d$ (bottom) distributions of one representative capsular contracture patient. Red contours represent peri-implant tissue.

**DLVH Comparison Between Capsular Contracture and Control Groups**

Three-dimensional surface DLVHs and corresponding two-dimensional iso-volume contour projections for peri-implant tissue are shown in Figure 2 for a representative capsular contracture patient and a matched control patient. The z-axis of the 3D DLVH represents the normalized cumulative volume receiving at least a given physical dose (x-axis) and $LET_d$ (y-axis). Each point on the surface thus reflects the fractional peri-implant volume exposed to a specific joint dose–$LET_d$ condition.

Compared with the matched control, the capsular contracture patient exhibited a substantially larger peri-implant volume exposed to high $LET_d$ (>5 keV/μm), as highlighted by the yellow arrows in the 3D surface plots (Figure 2a,c). This difference was more clearly visualized in the 2D iso-volume contour plots (Figure 2b,d), where the iso-volume lines for the capsular contracture patient were shifted toward higher $LET_d$ values across a broad range of doses. These shifts indicate increased overlap of moderate-to-high physical dose with elevated $LET_d$ in the peri-implant tissue.

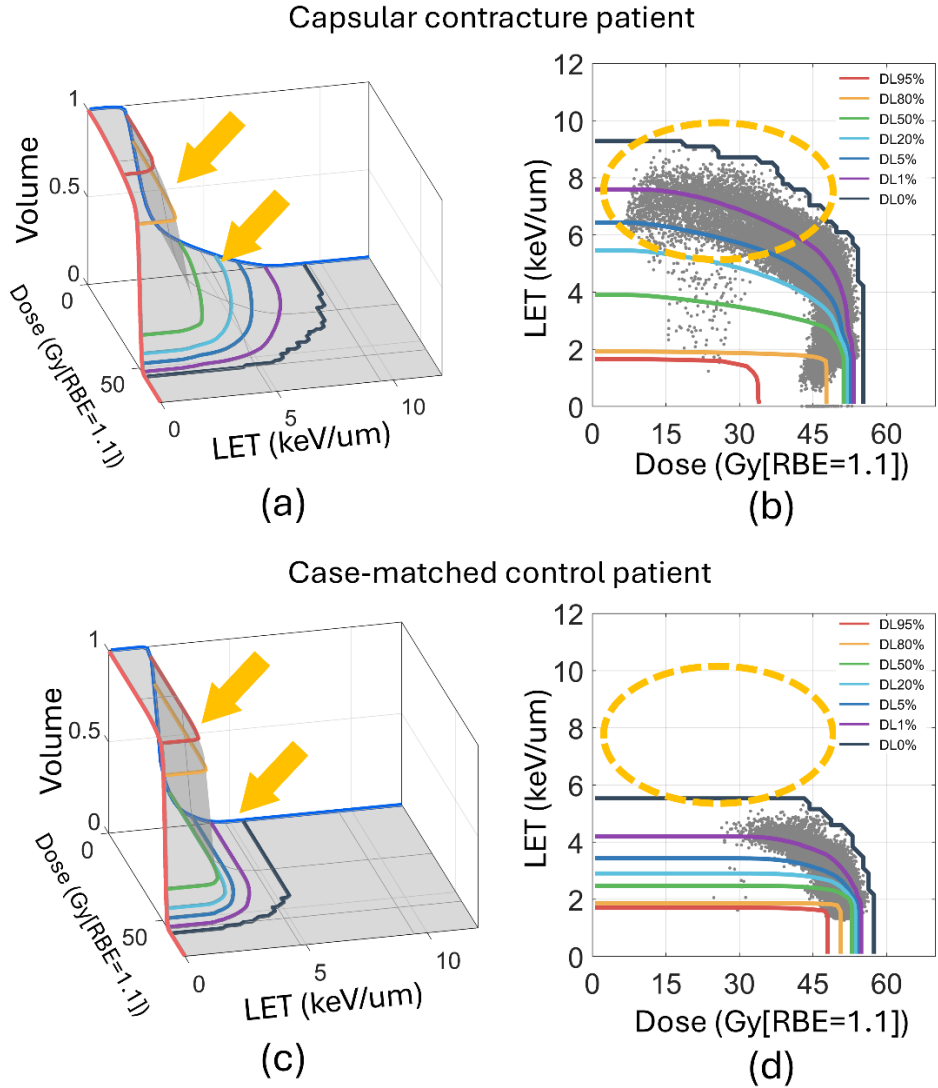

**Figure 2**. 3D surface plot and 2D iso-volume contour plot of Dose-LET volume histogram (DLVH) of peri-implant tissue for one AE patient and one matched control patient. Iso-volume lines of DL0%, 1%, 5%, 20%, 50%, 80%, and 95% were displayed. The DLVH index, *V(d, l)*, was defined as *V*(% for normalized volume) of the selected structure with a dose of at least *d* Gy and an $LET_d$ of at least *l* keV/μm. The shaded grey in 3D plot indicates the DLVH surface. Each gray dot in the 2D iso-volume contour plot corresponds to each voxel in the selected structure, mapping based on the physical dose and LET of the corresponding voxel. Yellow arrows in 3D plots and yellow arrows dashed circles in the 2D iso-volume contour plot highlight the difference of DLVHs between the capsular contracture patient and the matched control patient.

**Identification of DLVH Indices Associated with Capsular Contracture**

To systematically identify dosimetric features associated with capsular contracture, GLMER analyses were performed across all 4,096 possible combinations of dose and $LET_d$. The resulting regression coefficient $p$-values were visualized as a two-dimensional $p$-value map in the dose–$LET_d$ plane (Figure 3a). Distinct regions with statistically significant associations ($p < 0.01$) emerged, indicating specific joint dose–$LET_d$ conditions linked to capsular contracture risk.

Five candidate DLVH indices were initially selected from within these low–$p$-value regions: (55.8 Gy[RBE=1.1], 2.2 keV/μm), (50.3 Gy[RBE=1.1], 5.4 keV/μm), (43.8 Gy[RBE=1.1], 7.5 keV/μm), (36.0 Gy[RBE=1.1], 8.3 keV/μm), and (32.8 Gy[RBE=1.1], 0.9 keV/μm). Spearman correlation analysis was subsequently performed to evaluate redundancy among these indices. Correlation coefficient maps revealed strong intra-region correlations, indicating that multiple candidate indices represented similar dosimetric patterns (Figure 3b–d).

Based on these findings, two positively correlated indices, V(55.8 Gy[RBE=1.1], 2.2 keV/μm) and V(50.3 Gy[RBE=1.1], 5.4 keV/μm), were selected to represent the high-dose/moderate-$LET_d$ and moderate-dose/high-LETd regions, respectively. A third index, V(32.8 Gy[RBE=1.1], 0.9 keV/μm), exhibited negative correlation with both and was retained as a complementary feature representing favorable low-dose/low-LETd exposure.

Box plots comparing these three DLVH indices between the capsular contracture and control groups are shown in Figure 3e–g. Volumes corresponding to V(55.8 Gy[RBE=1.1], 2.2 keV/μm) and V(50.3 Gy[RBE=1.1], 5.4 keV/μm) were significantly higher in the capsular contracture group than in controls ($p = 0.0068$ and $p = 0.0054$, respectively), with near-zero volumes observed in most control patients. Conversely, V(32.8 Gy[RBE=1.1], 0.9 keV/μm) was significantly higher in the control group ($p = 0.0020$), indicating a protective association. Together, these findings demonstrate that capsular contracture is associated with increased peri-implant volume exposed to unfavorable dose-$LET_d$ combinations and reduced volume in favorable low-dose/low-$LET_d$ regions.

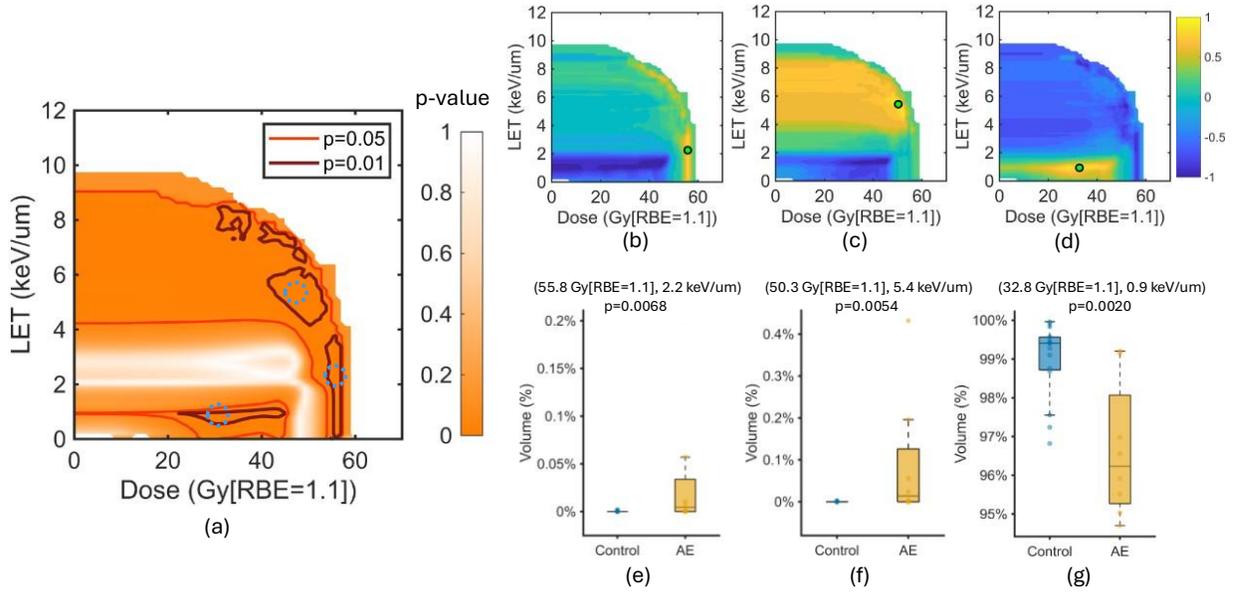

**Figure 3**. (a) Regression coefficient *p*-value color map derived from the generalized linear mixed-effects regression (GLMER) models for all DLVH indices. Iso-*p*-value lines of 0.01 and 0.05 were contoured in the color map. The DLVH indices correlated with capsular contracture are larger *V(55.8 Gy[RBE=1.1], 2.2 keV/μm)* and larger *V(50.3 Gy[RBE=1.1], 5.4 keV/μm)* (blue circles) and smaller *V(32.8 Gy[RBE=1.1], 0.9 keV/μm)* (green circles). (b-d) Spearman's coefficient map of the DLVH indices, which were indicated by green dots. *V(55.8 Gy[RBE=1.1], 2.2 keV/μm)* and *V(50.3 Gy[RBE=1.1], 5.4 keV/μm)* show relatively low correlation. But they both have negative correlation to *V(32.8 Gy[RBE=1.1], 0.9 keV/μm)*. Spearman's coefficient map of the other two redundant DLVH indices are present in SI. (e-g) Box plot of 3 significantly independent DLVH indices, with *p=0.0068, 0.0054* and *0.0020*, respectively.

**Derivation of Dose–LET Volume Constraints and NTCP Model Validation**

Receiver operating characteristic (ROC) analyses were performed for each of the three selected DLVH indices to derive clinically interpretable DLVCs (Figure 4a). The area under ROC curve (AUROC) values for V(55.8 Gy[RBE=1.1], 2.2 keV/μm) and V(50.3 Gy[RBE=1.1], 5.4 keV/μm) were 0.789 and 0.781, respectively, yielding optimal volume thresholds of <0.0033% and <0.0017%. These results indicate that even very small peri-implant volumes exposed to high-dose/moderate-LETd or moderate-dose/high-LETd conditions are associated with increased capsular contracture risk.

For V(32.8 Gy[RBE=1.1], 0.9 keV/μm), the AUROC was 0.898, and the optimal constraint was >96.98%, suggesting that maintaining the majority of peri-implant tissue within low-dose/low-LETd regions is protective. Classification accuracies for the three individual DLVCs were 87.5%, 83.3%, and 87.5%, respectively, demonstrating strong discriminative performance even when used independently.

To evaluate the combined predictive value of these DLVCs, a multivariable NTCP model was constructed using a support vector machine and validated via leave-one-out cross-validation. The resulting ROC curve is shown in Figure 4b, with an AUROC of 0.867. Individual NTCP predictions and true clinical outcomes are presented in Figure 4c. Using an NTCP threshold of 0.5, the model achieved an overall accuracy of 91.7%, sensitivity of 87.5%, and specificity of 93.8%. These results confirm that the derived DLVCs are robust, complementary predictors of capsular contracture risk and provide a strong foundation for biologically informed treatment planning optimization.

ROC performance of single DLVH index

| DLVH index | AUROC | Volume constrains | Accuracy | Sensitivity | specificity |
|---|---|---|---|---|---|
| (55.8 Gy[RBE=1.1], 2.2 keV) | 0.789 | 0.0033% | 87.5% | 62.5% | 100% |
| (50.3 Gy[RBE=1.1], 5.4 keV) | 0.781 | 0.0017% | 83.3% | 62.5% | 93.75% |
| (32.8 Gy[RBE=1.1], 0.9 keV) | 0.898 | 96.98% | 87.5% | 75% | 93.75% |

(a)

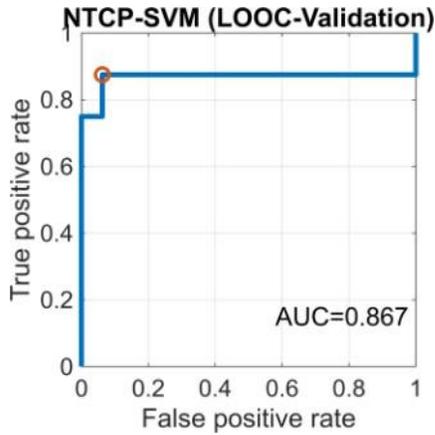
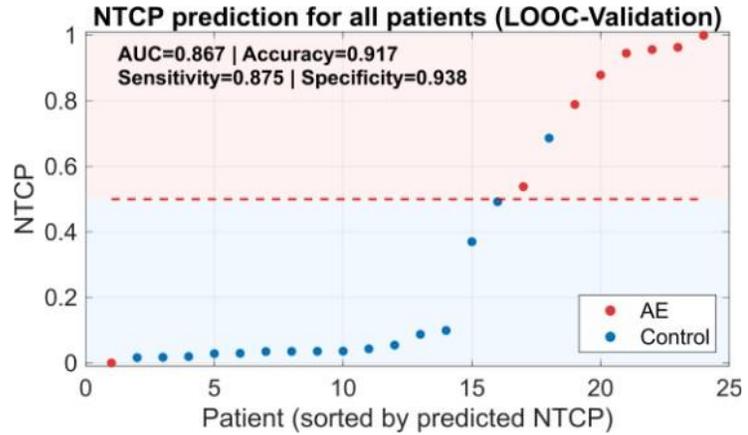

(b)  (c)

**Figure 4.** Dose-LET volume constrains extraction and performance of the SVM-based NTCP model. (a) The ROC performance of the three DLVH indices to predict the capsular contracture and volume constraints for the best performance. (b) ROC of the SVM-based NTCP model for the leave one out cross validation (LOOC-validation) (c) NTCP prediction for all patients for the LOOC-validation. The true patient outcome is indicated by the dot color. The background color and red dash line indicates the NTCP score below or above 0.5. The AUC, accuracy, sensitivity and specificity of the SVM-based NTCP model are 0.867, 0.917, 0.875 and 0.938, respectively.

**Discussion**

In this retrospective matched case–control study, we applied a DLVH-based mixed-effects regression framework to identify $LET_d$-related dosimetric features associated with capsular contracture following proton PMRT. Our findings demonstrate that capsular contracture is significantly associated with the synergistic interplay between physical dose and $LET_d$ within peri-implant tissue, rather than with either parameter alone. By integrating dose and $LET_d$ information at the volume level, we derived clinically interpretable dose-LET volume constraints (DLVCs), which were subsequently validated using an SVM-based NTCP model. Collectively, these results provide actionable dosimetric guidance for treatment planning optimization aimed at mitigating reconstructive complications after proton PMRT.

The study cohort consisted of 24 patients (8 with capsular contracture and 16 matched controls), with excellent balance across demographic, clinical, pathological, and surgical variables achieved through rigorous matching. This design minimized confounding effects and allowed focused investigation of dosimetric contributors to capsular contracture. The use of a GLMER model further accounted for within-set dependence and residual heterogeneity across matched sets, strengthening the robustness of the identified associations.

The DLVH *p*-value map revealed three distinct dose–$LET_d$ regions associated with capsular contracture risk. Two regions corresponded to high-dose/moderate-$LET_d$ and moderate-dose/high-$LET_d$ exposure, both of which were positively associated with capsular contracture. A third region, characterized by low-dose/low-$LET_d$, demonstrated a protective association, indicating that favorable peri-implant dosimetry is achieved by maintaining a large volume of tissue within this region. These findings align with emerging clinical evidence that spatial $LET_d$ heterogeneity, particularly near distal beam edges, can meaningfully influence normal tissue toxicity in proton therapy.

After eliminating redundant features through Spearman correlation analysis, three independent DLVH indices were identified as most predictive of capsular contracture The corresponding DLVCs, $DLVC_1$ (V(55.8 Gy, 2.2 keV/μm) < 0.0033%) and $DLVC_2$ (V(50.3 Gy, 5.4 keV/μm) < 0.0017%), highlight the importance of limiting even very small volumes of peri-implant tissue exposed to unfavorable dose-$LET_d$ combinations. Notably, the strong association observed for $DLVC_2$ indicates that relatively modest $LET_d$ values can contribute to toxicity when coupled with high physical dose, underscoring limitations of the conventional constant RBE = 1.1 assumption. $DLVC_3$ (V(32.8 Gy, 0.9 keV/μm) > 96.98%) highlights the importance of maintaining sufficient volume in the low-dose/low- $LET_d$ region to reduce the risk of adverse events. Our findings align with previous studies demonstrating associations between elevated $LET_d$ and late normal tissue complications, including rib fractures (28), late-phase pulmonary toxicity (31), brain necrosis (32-36), mandibular osteoradionecrosis (25), and rectal bleeding (37). Collectively, these results underscore $LET_d$ as a meaningful predictor of normal tissue toxicity and support the incorporation of LET-related biological effects into treatment planning to reduce late toxicities such as capsular contracture(38).

Validation using a multivariable SVM-based NTCP model demonstrated strong predictive performance (AUROC = 0.867, accuracy = 91.7%), confirming the clinical relevance of the extracted DLVCs. These results suggest that DLVCs can serve not only as post hoc risk stratification tools but also as prospective planning objectives.

A key strength of this work lies in the use of DLVH indices as outcome predictors. Compared with traditional dose- or $LET_d$-based metrics, the DLVH framework captures (1) volumetric effects, (2) the spatial co-occurrence of dose and $LET_d$, (3) phenomenologically derived dose-LETd relationships without reliance on uncertain RBE models, and (4) organ-level $LET_d$ integrity without violating statistical independence assumptions inherent to voxel-wise analyses. This approach provides a biologically informed yet clinically tractable methodology for toxicity modeling in proton therapy.

From a treatment planning perspective, current optimization workflows penalize dose or $LET_d$ violations independently, which fails to reflect their synergistic biological effects. The DLVCs identified in this study are directly compatible with emerging DLVC-based robust optimization (DLVCRO) strategies(48), which integrate dose, $LET_d$, and volume simultaneously into the optimization process. Prior work has demonstrated that DLVCRO can outperform conventional dose-volume- or LET-volume-based optimization in other disease sites, and our findings support its extension to breast proton therapy to reduce reconstructive toxicity.

Several limitations warrant consideration. Despite originating from one of the largest proton PMRT programs worldwide, however, the inherently low incidence of capsular contracture limited the number of events available for analysis, thereby constraining statistical power and potentially affecting generalizability. To address this limitation and ensure analytic rigor, our study focused specifically on clinically significant capsular contracture, and the resulting dose–LET volume constraints (DLVCs) were intentionally designed to prevent the most severe and clinically meaningful complications. These conservative thresholds may also provide secondary benefits by mitigating milder cosmetic sequelae, potentially preserving breast symmetry and patient-reported outcomes even among patients who do not develop overt contracture. Nevertheless, multi-institutional validation with larger cohorts will be essential to confirm the robustness of these DLVCs. Additionally, as with DVH-based analyses, DLVH inherently sacrifices spatial information, and regions of extreme dose-$LET_d$ exposure often involve very small volumes, increasing susceptibility to statistical uncertainty. Future work incorporating spatially resolved or biomechanically informed models may further refine risk prediction. Finally, while capsular contracture represents an important source of morbidity following breast reconstruction, infectious complications also contribute significantly to patient burden. Further studies are needed to investigate strategies for reducing these risks following both photon and proton therapy.

## Conclusion

In this retrospective matched case–control study, we developed and applied a DLVH-based analytical framework to investigate the relationship between combined dose-$LET_d$ distributions and capsular contracture following proton PMRT. Three independent dosimetric features within peri-implant tissue were identified as significant predictors of capsular contracture risk, leading to the derivation of clinically interpretable dose-LET volume constraints: V(55.8 Gy[RBE=1.1], 2.2 keV/μm) < 0.0033%, V(50.3 Gy[RBE=1.1], 5.4 keV/μm) < 0.0017%, and V(32.8 Gy[RBE=1.1], 0.9 keV/μm) > 96.98%. These constraints were validated using an SVM-based NTCP model, which demonstrated strong discriminative performance (AUROC = 0.867).

Our findings indicate that capsular contracture is driven by the synergistic interaction between physical dose and LETd, rather than by either parameter alone, highlighting limitations of the constant RBE paradigm in proton therapy. The proposed DLVCs provide actionable guidance for treatment planning and represent a critical step toward biologically informed optimization strategies. Integration of these constraints into future DLVC-based robust optimization workflows may help reduce reconstructive toxicity while preserving the dosimetric advantages of proton PMRT.